# Neutron scattering by magnetic octupoles of a quantum liquid

**Nicolas Gauthier[1,2], Victor Porée[1], Sylvain Petit[3,*], Vladimir Pomjakushin[1,*], Elsa Lhotel[4], Tom Fennell[1] & Romain Sibille[1,*]**

[1] Laboratory for Neutron Scattering and Imaging, Paul Scherrer Institut, 5232 Villigen PSI, Switzerland.

[2] Stanford Institute for Materials and Energy Science, SLAC National Accelerator Laboratory and Stanford University, Menlo Park, California 94025, USA.

[3] LLB, CEA, CNRS, Université Paris-Saclay, CEA Saclay, 91191 Gif-sur-Yvette, France.

[4] Institut Néel, CNRS–Université Joseph Fourier, 38042 Grenoble, France.

[*] email: romain.sibille@psi.ch ; sylvain.petit@cea.fr ; vladimir.pomjakushin@psi.ch

**Neutron scattering is a powerful tool to study magnetic structures and dynamics, benefiting from a precisely established theoretical framework. The neutron dipole moment interacts with electrons in materials via their magnetic field, which can have spin and orbital origins. Yet in most experimentally studied cases the individual degrees of freedom are well described within the dipole approximation, sometimes accompanied by further terms of a multipolar expansion that usually act as minor corrections to the dipole form factor. Here we report a unique example of neutrons diffracted mainly by magnetic octupoles. This unusual situation arises in a quantum spin ice where the electronic wavefunction becomes essentially octupolar under the effect of correlations. The discovery of such a new type of quantum spin liquid that comes with a specific experimental signature in neutron scattering is remarkable, because these topical states of matter are notoriously difficult to detect.**

## ON MAGNETIC NEUTRON SCATTERING

Neutrons, thanks to their spin, are employed to discern materials' magnetic properties. They are sensitive to the magnetic field produced by unpaired electrons, which can have various symmetries and properties depending on the particular atom and its crystal field environment. The magnetization density originates from both the spin and orbital distributions of open shell electrons and can be expanded in multipoles with the use of spherical harmonics. In a vast number of cases, neutron scattering results are well accounted for by considering the magnetic dipole moment of the atom – a parity-even tensor of rank 1 (axial vector). The tensors of higher odd-rank $K$ are conventional magnetic multipoles such as the octupole ($K$=3) and the triakontadipole ($K$=5). However, their contribution to the neutron scattering form factor is usually marginal in comparison to the dipole moment.

The interaction of the magnetic multipole degrees of freedom with the neutron spin are described by the neutron-electron interaction operator $\mathbf{Q}=\exp(i\mathbf{qr})(\mathbf{s}-ih/q^2\ [\mathbf{q}\times\mathbf{p}])$ [1]. This operator can be expressed using spherical Bessel functions $j_n\ (q)$ in powers of $(\mathbf{qr})^m$ (m=0,1,2...), where **q** is the neutron scattering wavevector, and **s**, **r** and **p** are the spin, position and momentum of the electron. The first two leading terms in this expansion give the so-called 'dipole approximation'. The dominating contribution to the neutron scattering is given by the conventional radial integral $\langle j_0\ (q)\rangle$, which has maximum at $q$=0. Higher terms in the expansion of **Q** contain the contribution of the conventional magnetic multipole moments. The calculations of these higher-order contributions to the scattered neutron intensity is mathematically quite involved and requires the use of spherical tensors and Racah tensor algebra with the detailed procedures given in [1]. Despite this complexity, we can identify two main characteristics expected from the conventional magnetic multipoles based on the neutron-electron interaction. First, conventional magnetic multipoles give significantly smaller contribution to neutron scattering than the dipole one. Second, their form-factor is zero at $q$=0 with a maximum at relatively high $q$, and is also anisotropic. The expected signatures of magnetic multipoles are therefore a weak anisotropic signal at high q, making their experimental detection a real challenge.

Multipoles that are observable in neutron scattering must be odd under time-reversal symmetry. This includes the conventional magnetic multipoles, which are parity-even multipoles and are the main topic of this article. We note, however, that parity-odd multipoles can also exist and be observed if the atomic wavefunction does not have a well-defined parity [2-4]. These parity-odd multipoles are fundamentally different from the conventional (parity-even) magnetic multipoles. For example, the first order parity-odd multipoles are called anapoles, or toroidal moments, which are the cross products of spin s or orbital l angular momentum with the electron position **r**.

## MULTIPOLES IN CONDENSED MATTER RESEARCH

Although a vast majority of materials with strong electronic correlations can be well understood based on individual degrees of freedom described using the first term of the multipolar expansion, further terms are required to explain an increasing number of novel phenomena. Such multipole moments can in principle lead to the emergence of macroscopic orders that are sometimes called 'hidden' due to the challenge of determining their order parameter [5-6]. Multipoles in condensed matter correspond to anisotropic distributions of electric and magnetic charges around given points of the crystal structure – a situation that can arise at the atomic scale from spin-orbit coupling, such as for the multipoles proposed to explain a famously mysterious phase in the heavy-fermion material $URu_2Si_2$ [7], or at the scale of atomic clusters where the established correlations lead to the emergence of novel degrees of freedom, such as in the spin-liquid regime of $Gd_3Ga_5O_{12}$ [8].

As already noted, neutrons are also sensitive to odd-parity multipoles, and a number of studies have pointed to their role to explain phase transitions that break both space inversion and time reversal. This has been especially discussed in the context of magnetoelectric insulators [9-10], and in high-$T_C$ superconductors where magneto-electric quadrupoles were proposed as the order parameter of the transition appearing in the pseudogap region [11-12].

Well characterized examples of 'hidden' orders of (conventional) multipoles exist, such as in $NpO_2$ or $CeB_6$ and its substitutional alloys $Ce_{1-x}La_xB_6$ [5-6]. In neptunium dioxide [13], the primary order parameter is associated with magnetic octupoles that order around 25 K in a longitudinal structure defined by three wavevectors. This structure of ordered octupoles was established, indirectly, using resonant X-ray diffraction through the measurement of a parasitic order of electric quadrupoles having the same structure [14,15]. In cerium hexaboride, the first correlated phase entered upon cooling in zero field below $T_Q$ = 3.4 K is an antiferroquadrupolar order, which is followed by an antiferromagnetic order of dipoles at $T_N$ = 2.3 K [16,17]. The dipole-dipole nature of the intersite magnetic interactions make the associated correlations more resistant against disorder than for the electric quadrupoles, so that $T_Q$ decreases faster than $T_N$ upon diluting the cerium lattice with lanthanum [18]. At some doping level, these phase transitions intersect and a new phase pocket appears, characterized by an antiferromagnetic ordering of octupoles that was measured directly using resonant X-ray diffraction [19].

The contribution of parasitic magnetic octupoles to the total magnetic scattering intensity – of mainly dipole origin, is well known in materials such as elemental holmium for instance [20]. However, the experimental results for $Ce_{0.7}La_{0.3}B_6$ presented in ref. [21], where magnetic multipoles are the primary order parameter, is to the best of our knowledge a unique example in terms of ordered magnetic multipoles scattered by neutrons. Only three independent magnetic Bragg peaks were detected, but the fact that the intensity at $q$=6 Å$^{-1}$ is larger than at $q$=1.3 Å$^{-1}$ has led the authors to argue that these have octupolar origin. A later theoretical study [22] qualitatively confirms that the q-dependence of the observed peaks agrees with the calculated octupolar neutron magnetic form factor of cerium in this material.

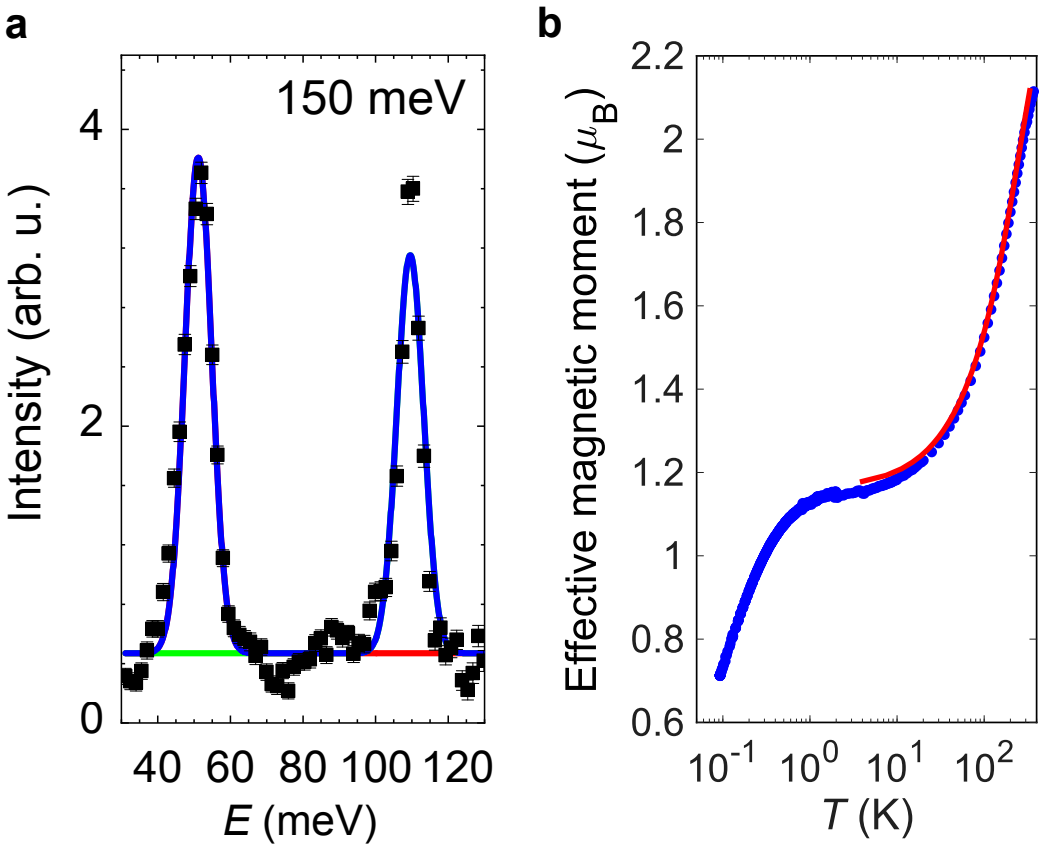

**Figure 1**
Inelastic neutron scattering (INS) data of $Ce_2Sn_2O_7$ probing the crystal-electric field levels within the ground multiplet $^2F_{5/2}$ of $Ce^{3+}$ (**a**) [29]. The bulk magnetic susceptibility $\chi$ [28], shown in blue as the effective magnetic moment $\mu_{eff} \propto \sqrt{\chi T}$ as a function of temperature (**b**) reveals three regimes. At high (T › 100 K) temperature $\mu_{eff}$ decreases due to crystal-field effects – a regime that is well reproduced by the calculation from the fit of the INS data [29]. At moderate (1 K ‹ T ‹ 10 K) temperatures, a plateau of ~1.2 $\mu_B$ corresponds to the dipole moment of the ground state 'dipole-octupole' doublet. At low (T ‹ 1 K) temperature, $\mu_{eff}$ decreases further under the effect of dominant octupole–octupole correlations, making the octupole moment strengthen at the expense of the dipole one, as shown in Figure 2.

A constant to all materials discussed so far is the notion of order associated with multipoles. Comparatively, far less attention has been given to the effect of multipolar fluctuations in the absence of multipolar order. We can mention, though, that a number of recent theoretical studies point to the relevance of multipole fluctuations in unconventional superconductivity [23-26], where it is proposed to induce anisotropic pairing mechanisms [27].

A central result of the present work is to demonstrate the existence of a phase of correlated yet fluctuating multipoles, in the absence of long-range order down to the lowest temperatures. This discovery is based on a number of signatures indicating the development of a correlated ground state below 1 Kelvin, corroborated by neutron scattering results consistent with the formation of a fluid-like state of magnetic octupoles. The intensity distribution is weighted to large scattering vectors, which indicates that the correlated degrees of freedom have a more complex magnetization density than that typical of magnetic dipoles in a spin liquid. Keeping the broader relevance and implications of these findings for the last section of this article, we can already mention here that, in the context of neutron scattering, $Ce_2Sn_2O_7$ already stands out as it appears to be the second experimental evidence of neutrons scattered by non-parasitic higher-order magnetic multipoles.

## MAGNETIC MULTIPOLES IN CERIUM STANNATE

We first define the context of the present study in terms of uncorrelated degrees of freedom in the material of interest, $Ce_2Sn_2O_7$ [28-29]. Trivalent cerium has one f electron, with spin *S*=1/2 and orbital angular momentum *L*=3 mixed into a *J*=5/2 ground multiplet by spin-orbit coupling, which splits into three Kramers doublets in the crystal-electric field. Transitions within this ground multiplet are easily seen in inelastic neutron scattering measurements using an incident energy $E_i$ of 150 meV, at energy transfers $E \approx 51$ and 110 meV (Figure 1**a**) [29]. Intermultiplet transitions to the *J*=7/2 multiplet are also visible from measurements taken with higher $E_i$, but their overlap in this data make them rather ill-defined experimentally. However, the transitions within the ground multiplet are sufficient to refine the parameters of the crystal-field Hamiltonian, and these can reproduce the data at higher energy transfers as well as the bulk susceptibility at high temperature. The latter is represented in Figure 1**b**, where the effective magnetic dipole moment $\mu_{eff} \propto \sqrt{\chi T}$ is plotted as a function of temperature. This quantity decreases upon depopulating excited crystal-field levels when cooling, to reach an approximate plateau of ~1.2 $\mu_B$ in the range from 1 to 10 K. This value corresponds to the dipole moment calculated from the wavefunction of the ground doublet only.

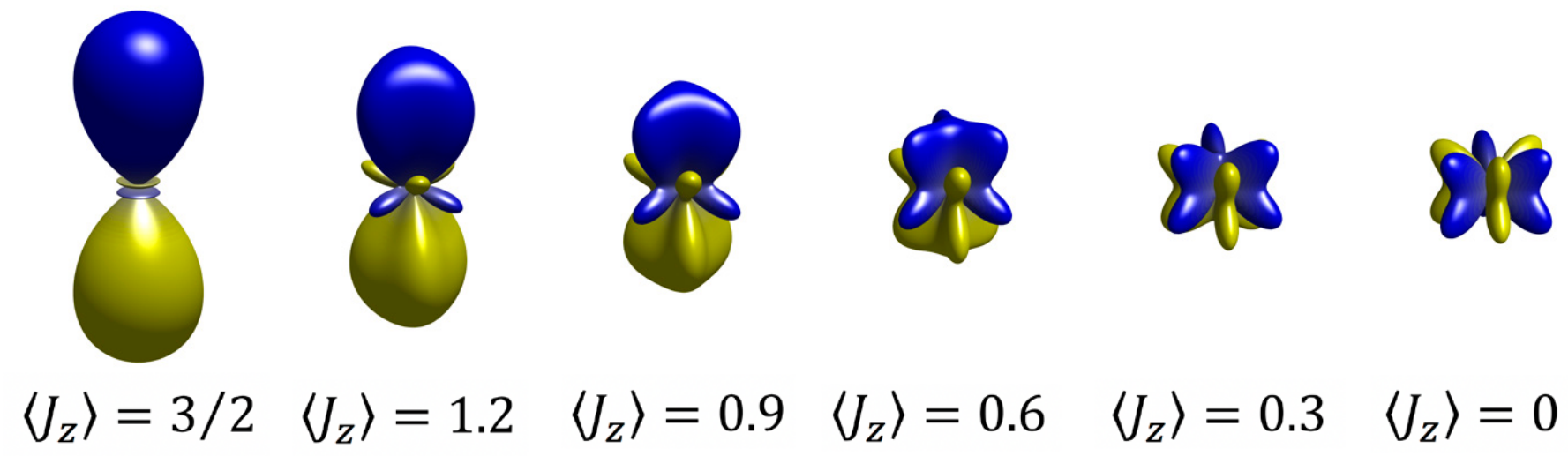

**Figure 2**
Magnetic charge density calculated from the type of ground state wavefunction of $Ce^{3+}$ determined from the fit of the neutron data in Figure 1, i.e. $|\pm\rangle = A|\pm 3/2\rangle \pm B|\mp 3/2\rangle$. The values of $J_z$ correspond to different values of the *A* and *B* coefficients.

The essential result of the above crystal-field analysis is that the wavefunction of the ground state Kramers doublet is of the general form $|\pm\rangle = A|\pm 3/2\rangle \pm B|\mp 3/2\rangle$, where A and B are real. Any linear combination of these $|m_{J_z} = \pm 3/2\rangle$ states is by definition an eigenstate of the Hamiltonian. The wavefunction corresponds to a so-called 'dipole-octupole' doublet [30-31]. Importantly, we point out that the moment sizes respectively associated with the dipole and octupole operators directly depend on the values of the *A* and *B* coefficients. For instance, the situation where *A*=1 and *B*=0 leads to a full dipole moment $\langle J_z \rangle = 3/2$ and a zero octupolar moment, while A=B gives $\langle J_z \rangle = 0$ but a net octupolar moment (Figure 2). A key idea of the present work on $Ce_2Sn_2O_7$ is that the further decrease of dipole moment, observed in $\mu_{eff}$ when cooling down in the correlated regime below 1 K, is due to dominant octupole–octupole couplings, caus-

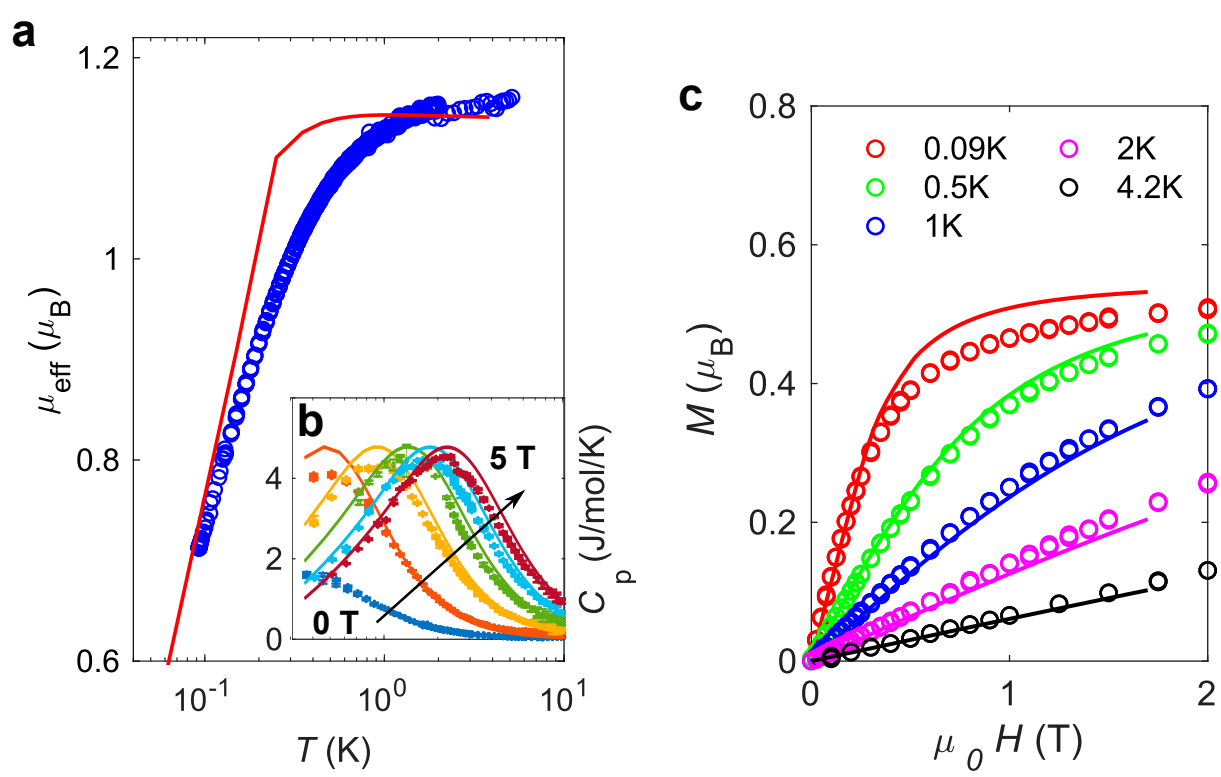

**Figure 3**

Effective magnetic moment $\mu_{eff} \propto \sqrt{\chi T}$ (**a**) and heat capacity (**b**) as a function of temperature in the correlated regime of $Ce_2Sn_2O_7$ below 1 Kelvin [28-29]. The magnetization curves as a function of field are shown in panel (**c**). All experimental data are shown as open or close circles and were used to fit the dipole–dipole $\mathcal{J}^{zz}$ and octupole–octupole $\mathcal{J}^{yy}$ exchange parameters using the relevant Hamiltonian for 'dipole-octupole' doublets on the pyrochlore lattice.

ing the octupole moment to strengthen at the expense of the dipole one. In other words, dominant octupole–octupole interactions mix the otherwise degenerate $|m_{J_z} = \pm 3/2\rangle$ states to form new split eigenstates – the driving force being to minimize the energy of the system due to different magnetic dipole and octupole moment sizes.

## SIGNATURES OF CORRELATIONS IN MACROSCOPIC MEASUREMENTS

Our investigations of $Ce_2Sn_2O_7$ started with measurements of the bulk magnetization and heat capacity down to very low temperature [28]. As already exemplified with the plot of $\mu_{eff}$ shown in Figure 1**b**, signatures of a correlated state appear in this data below about 1 Kelvin. At these temperatures, the ground state doublet is thermally well isolated from excited crystal-field levels, and therefore is sufficient as a minimal low-energy description of the degrees of freedom. In the correlated regime, where the magnetic susceptibility increases slower than expected for a simple paramagnet, a hump is also observed in the heat capacity (Figure 3), thus further hinting at cooperative phenomena setting-in below 1 Kelvin.

The magnetization curves are also instructive, as the powder-averaged saturation at high field occurs at roughly half the value of the ground-doublet dipole moment, which is expected for Ising moments on a pyrochlore lattice due to the important noncollinear local anisotropy. The Ising anisotropy of the dipoles is also corroborated by calculations using the wavefunction determined from the analysis of the inelastic neutron scattering results. It is interesting to note that Ising moments of ~1.2 $\mu_B$ on the pyrochlore lattice of $Ce_2Sn_2O_7$ translate into classical dipole–dipole couplings of about 0.025 K, which is small compared to the scale of the dominant interactions. This simple comparison indicates that the correlated state originates from quantum-mechanical exchange interactions [28]. All bulk measurements, as well as muon spin spectroscopy [28], exclude the presence of magnetic order in $Ce_2Sn_2O_7$ down to the lowest temperatures (0.02 K), and instead suggest a highly frustrated magnet.

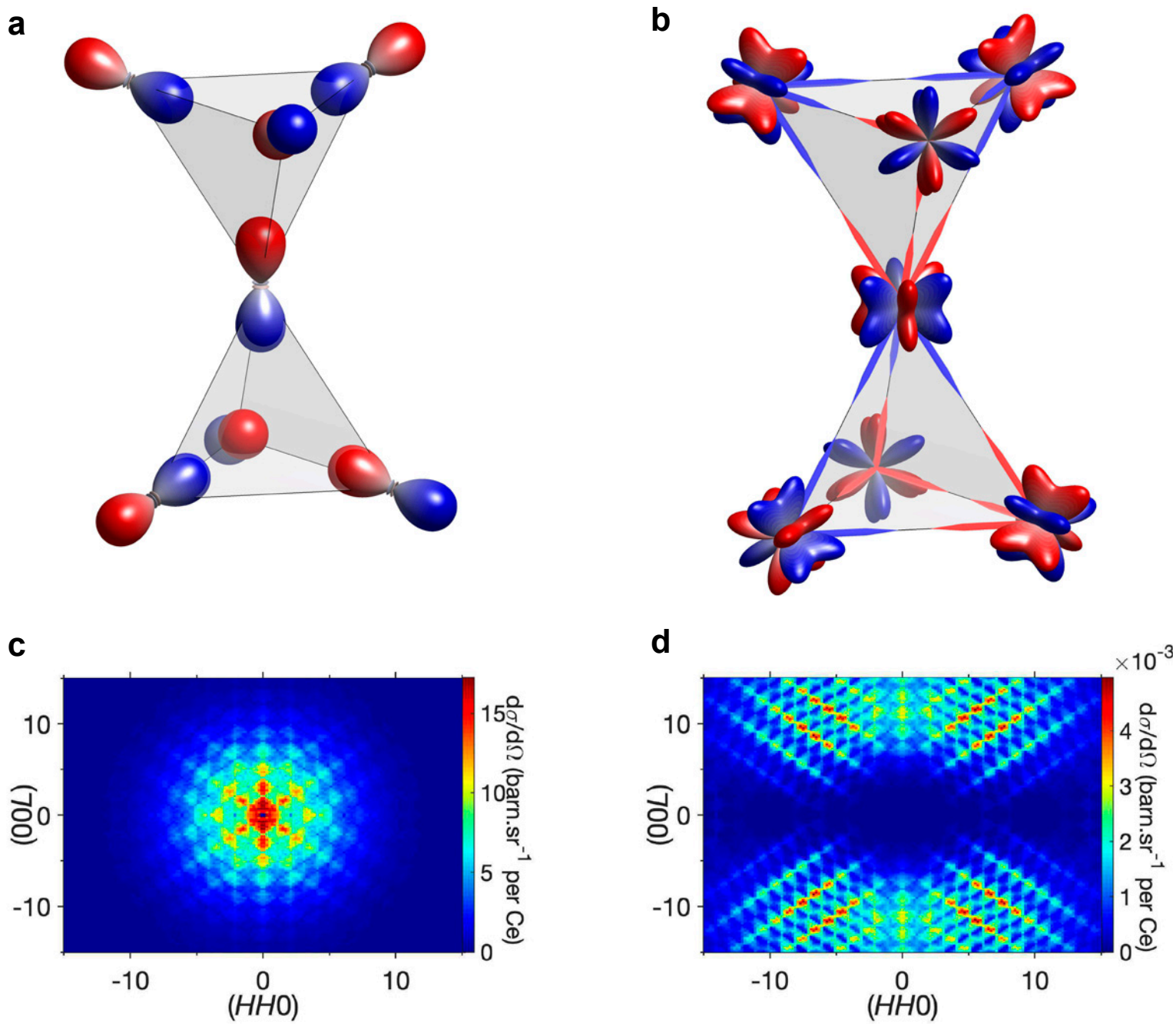

**Figure 4**
Magnetic dipoles respecting the '2-in-2-out' ice rule on each tetrahedron (**a**) and octupoles obeying the '2-plus-2-minus' rule (**b**), together with their respective neutron magnetic diffuse scattering patterns (**c** and **d**) calculated in the (HHL) plane of reciprocal space using Monte Carlo simulations [29]. Note that the spin ice pattern (panel **c**) is displayed over a much larger area of reciprocal space than usual, but the typical features can be discerned in the central region.

The set of bulk measurements presented in Figure 3 can be used to extract exchange constants using the relevant Hamiltonian for rare-earth pyrochlores with 'dipole–octupole' doublets, $\mathcal{H}_{DO} = \sum_{<ij>}[\mathcal{J}^{xx}\tau_i^x\tau_j^x + \mathcal{J}^{yy}\tau_i^y\tau_j^y + \mathcal{J}^{zz}\tau_i^z\tau_j^z + \mathcal{J}^{xz}(\tau_i^x\tau_j^z + \tau_i^z\tau_j^x)]$ [30-31]. The doublet is modelled by pseudo-spin $S = 1/2$ operators $\vec{\tau}_i = (\tau_i^x, \tau_i^y, \tau_i^z)$, where the components $\tau_i^x$ and $\tau_i^z$ transform like magnetic dipoles while $\tau_i^y$ behaves as an octupole moment. For the sake of simplicity and in order to avoid over-parametrizing the fit, we consider $\mathcal{J}^{xx} = \mathcal{J}^{xz} = 0$, which still captures the essential physics of octupolar phases. Using mean-field calculations, the bulk magnetic properties at low temperature are employed to extract values for $\mathcal{J}^{yy}$ and $\mathcal{J}^{zz}$ (see Figure 3) [29]. As already explained at a qualitative level, the drop of the effective magnetic moment below 1 K can be accounted for using a dominant octupole–octupole interaction $\mathcal{J}^{yy}$. Although the fit is equally good for $\mathcal{J}^{yy}$ =+0.48 ± 0.06 K or $\mathcal{J}^{yy} = -0.16$ ± 0.02 K (and a small but finite dipole–dipole coupling $\mathcal{J}^{zz}$ =+0.03 ± 0.01 K), only $\mathcal{J}^{yy} > 0$ corresponds to a frustrated arrangement of octupoles and is able to explain the absence of phase transition. In this case, the magnetic charge density of the octupoles on a tetrahedron is constrained by a '2-plus-2-minus' ice rule (Figure 4), leading to an extensively degenerate manifold of octupole ice configurations. In the present context, 'plus' and 'minus' replace the 'in' and 'out' states characterizing dipole Ising moments on the corner-sharing lattice of tetrahedra in spin ices, and instead designate the two possible local mean-values of the octupolar operator associated with $\tau_i^y$. We note that $\mathcal{J}^{yy} < 0$ would imply an ordered phase of octupoles with 'all-plus-all-minus' configurations on each tetrahedron, which is not frustrated and would lead to a phase transition.

## MAGNETIC OCTUPOLE NEUTRON SCATTERING IN CERIUM STANNATE

We now turn to the observation of magnetic neutron scattering from the octupole ice state expected from the analysis of the bulk measurements. Using the powder diffractometer HRPT at SINQ, later also confirmed using D20 at ILL, a weak diffuse signal appearing at high scattering vectors was observed in high-statistics difference data between 5 K and several lower temperatures ranging from 2 K to 0.05 K (Figure 5) [29]. The intensity distribution is zero at low scattering vectors $q < 4$ Å$^{-1}$ but grows and reaches a maximum around $q \sim 8$ Å$^{-1}$, which is characteristic of higher-order multipoles.

The diffuse scattering observed in $Ce_2Sn_2O_7$ has several important implications, even though its wave-vector dependence cannot be studied in details due to powder averaging. First, the simple existence of this signal is likely to be an additional and good reason to think that the system is governed by quantum exchange interactions. Second, the magnitude of the signal, presented in absolute units, is in very good agreement with that expected for a ground state based on octupole ice correlations, thus giving strong support to their existence in this material. To ascertain this, the experimental data in Figure 5 is presented together with the powder average of the diffuse neutron scattering calculated for an octupole ice phase. The signal measured at 0.05 K is about two thirds of the intensity of the zero-temperature calculation, the latter assuming a full octupolar moment and the generic wavefunction of $Ce^{3+}$ determined for $Ce_2Sn_2O_7$.

The octupole ice scattering is in general about several hundred times weaker than the value expected for spin-ice scattering, even considering the small dipole moment of $Ce_2Sn_2O_7$, which itself is much weaker than that of classical spin ices like $Ho_2Ti_2O_7$ [32]. Therefore, it remains a real challenge to measure such a weak signal, which was repeated multiple times and on two different instruments

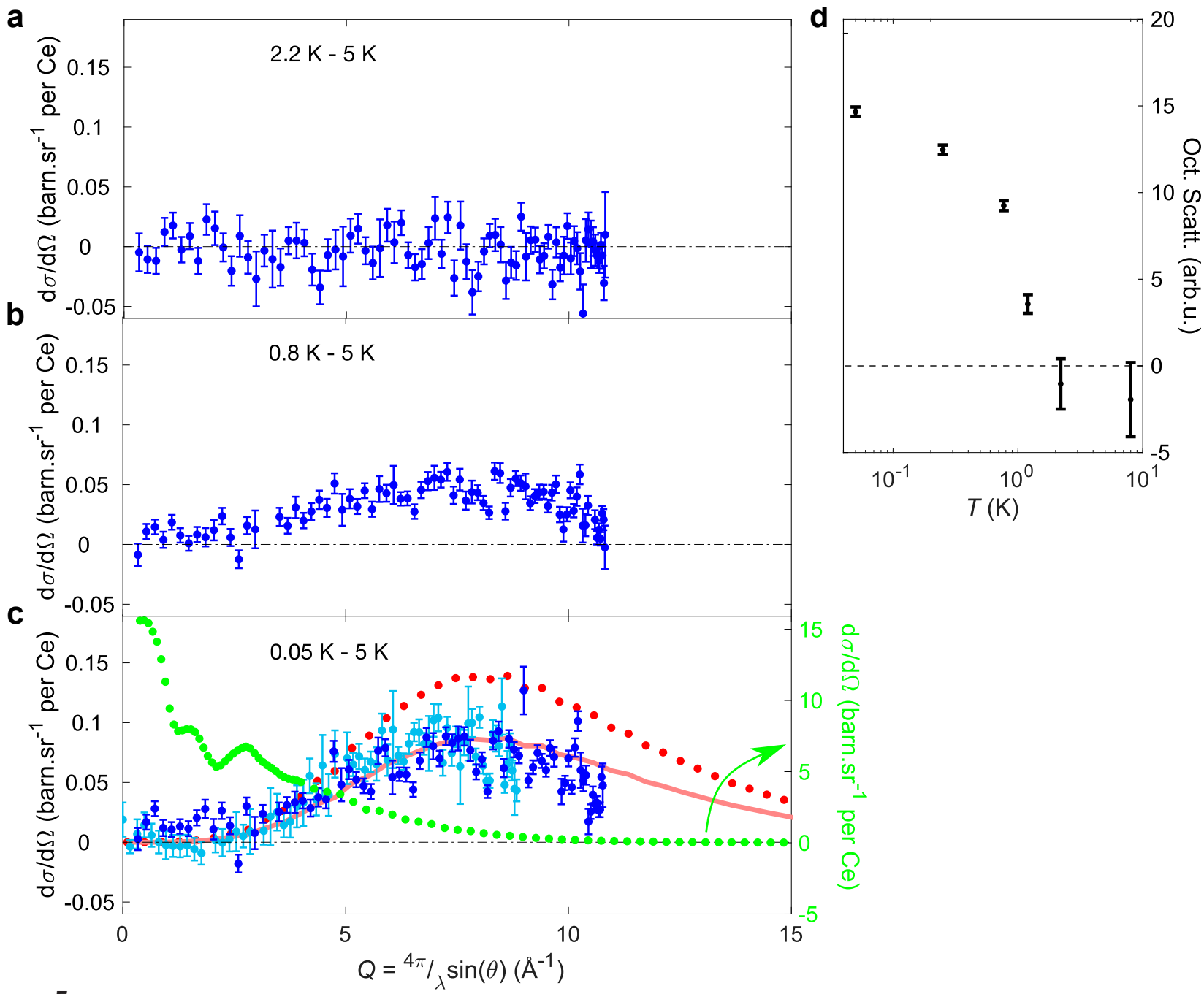

**Figure 5**

Diffuse octupolar scattering (blue points with error bars corresponding to ±1 standard error) obtained from the difference between neutron diffraction patterns measured at 5 K and at a lower temperature indicated on each panel. Measurements were performed on HRPT (λ = 1.15 Å, dark blue points on panels **a-c**) and D20 (λ = 1.37 Å, light blue points on panel **c**). Note the large scattering vectors required to observe scattering by magnetic octupoles. The increase of octupolar moment evaluated by the temperature dependence of the integrated diffuse scattering (**d**) matches with the drop of the dipole moment measured in bulk susceptibility (c.f. Figure 3**a**). The powder average of the diffuse scattering calculated for the octupole ice and spin ice configurations is shown respectively with red and green points in panel c, while the solid red line represents the same calculation for the octupole ice but scaled (× 0.625) onto the experimental data. Note the different scales used to display the octupolar (left scale) and dipolar (right scale) scattering in $Ce_2Sn_2O_7$.

in order to exclude spurious origins of the hump observed at high q. The available neutron flux is actually not the most important instrumental characteristic to improve the statistics here, but rather a combination of well understood instrument features, background, collimation, high resolution in order to distinguish the signal from the shifts of Bragg peaks due to thermal contraction in difference patterns, and optimal choice of wavelength in order to favour a large detector angular range for the region of interest in reciprocal space.

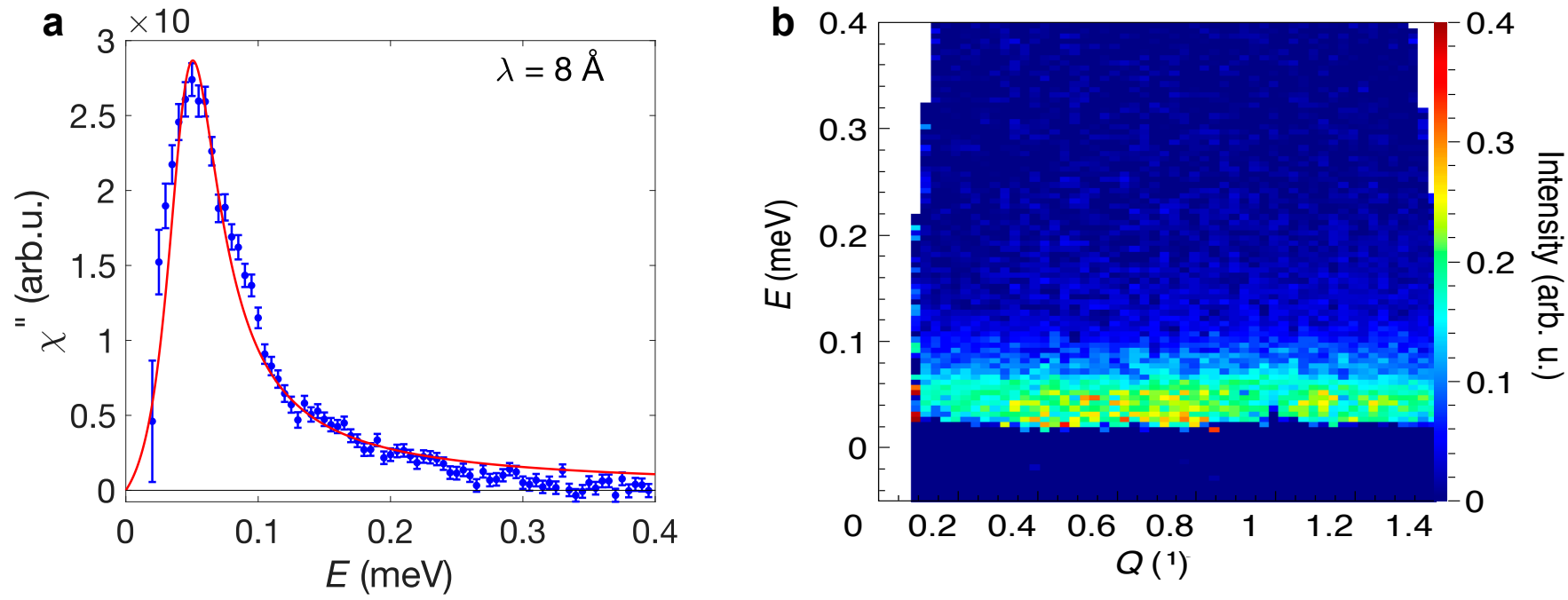

**Figure 6**
Imaginary part of the dynamic spin susceptibility (**a**, blue points with error bars corresponding to ±1 standard error). The difference map between low (correlated) and high (uncorrelated) temperatures (**b**), summarizes the wavevector dependence of the spin excitations as a function of energy, giving evidence for a continuum of fractionalized spin excitations. The phenomenological form used to fit the spectrum (red line) happens to capture the features expected from theory for spinon excitations in a quantum spin ice (onset, peak and extent) [33,37-39].

## AN OCTUPOLAR QUANTUM SPIN ICE

The exchange constants extracted from the set of bulk measurements place $Ce_2Sn_2O_7$ in the octupolar quantum spin ice regime [30-31] – a quantum liquid built up from a manifold of octupole ice configurations set by the dominant $\mathcal{J}^{yy}$ exchange and allowed to quantum fluctuate thanks to $\mathcal{J}^{zz}$ acting as a transverse perturbation.

In a (dipolar) quantum spin ice, excitations are expected to be of two types [33]. Gapped excitations, akin to spinons, correspond to defects created by single spin-flips in a '2-in-2-out' manifold – a quantum version of the magnetic monopoles of classical spin ice. Equivalently, in an octupolar quantum spin ice [30-31], we still expect such excitations to arise in the form of dipolar low-energy inelastic neutron scattering, because $\mathcal{J}^{zz}$ allows neutron-active transitions between the two states of the doublet split by $\mathcal{J}^{yy}$ [29]. Low-energy neutron spectroscopy data of $Ce_2Sn_2O_7$ indeed reveal the presence of low-energy excitations that are dipolar in nature (Figure 6) [29], and their continuous character matches expectations for the fractionalized spinon excitations of a quantum spin liquid. Gapless excitations of a quantum spin ice known as 'photon' excitations, however, are expected to follow an octupolar form factor in the present case. And, given the bandwidth expected for such excitations, in the μeV range, resolving their energy spectrum using neutron scattering appears impossible in view of the incident energies of the order of 50 meV required to reach far enough reciprocal space.

The continuum is peaked around 0.05 meV, which is approximately the dominant exchange interaction $\mathcal{J}^{yy}$, and extends up to at least 0.3 meV, which is significantly larger. Both of these characteristics are consistent with theoretical predictions for the spinon continuum of a quantum spin ice [33,37-39]. Moreover, at the low-energy edge of the continuum, a sharp increase in the density of states is observed, which is reminiscent of the threshold predicted for quantum spin ices due to the effect of the emergent photon on the production of spinons [39].

## COMPARISON WITH OTHER MATERIALS

Recent works have focused on a related material, $Ce_2Zr_2O_7$, with the advantage of large single crystals being available [34-35]. However, this comes at the price of a poor control of the oxygen stoichiometry due to the difficulty to maintain a clean atmosphere at the very elevated melting temperature reached during travelling-solvent crystal growth, combined with a problem of oxidation at ambient conditions inherent to the compound [34]. Oppositely, the trivalent oxidation state of cerium can be readily stabilized in $Ce_2Sn_2O_7$, taking advantage of a solid state oxydo-reductive reaction during which $Sn^0$ is oxidized to $Sn^{4+}$ while reducing $Ce^{4+}$ to the required $Ce^{3+}$ [36]. Moreover, the oxygen stoichiometry of the stannate can be precisely determined by thermogravimetric measurements under air by following the mass gain from the sample to its oxidation product, which leads to conclude for a perfect oxygen stoichiometry within a precision of less than 1 %. This contrasts with the 10 % of extra oxygen, and hence non-magnetic $Ce^{4+}$ impurities, in samples of cerium zirconate [34].

Despite these differences in the quality of the samples of $Ce_2Sn_2O_7$ and $Ce_2Zr_2O_7$, a number of measurements indicate that both materials are characterized by the same quantum spin liquid ground state. These are *i*) a crystal-field analysis indicating a 'dipole–octupole' ground state doublet in both cases [29,34-35], *ii*) heat capacity data that appear consistent for the two materials [28,34-35], *iii*) muon spin rotation experiments indicating that both of these isostructurals retain a dynamical magnetic ground state down to 0.02 K [28,35], and *iv*) low-energy neutron spectroscopy experiments providing evidence for continua of dipolar spin excitations characterized by the same energy and damping in $Ce_2Sn_2O_7$ [29] and $Ce_2Zr_2O_7$ [35]. For all these reasons, it appears reasonable to think that the octupolar quantum spin ice scenario might also explain the observations of quantum spin liquid dynamics in $Ce_2Zr_2O_7$.

To broaden the scope of the comparison to other materials as well, we can mention here that $Ce^{3+}$ pyrochlores hold a certain degree of uniqueness in the whole zoo of candidate quantum spin liquids. Indeed, they have smaller dipole moments compared to most candidates based on rare-earths [40], thus reducing the classical dipole–dipole forces to a negligible interaction, and their ground-state doublet is well isolated – thus clearing any doubts on possible quantum processes involving low-lying crystal-field levels [41]. The Kramers nature of the ground state doublet additionally avoids complexities related to transverse fields induced by the residual amounts of structural disorder always present in any real sample [42-45]. These characteristics, together with the cooperative behavior of the 'dipole-octupole' pseu-do-spins we have demonstrated and the excitation continuum we have measured, give these materials the promise to become excellent proven examples of a three-dimensional quantum spin liquid.

## CONCLUSIONS AND OUTLOOK

With this experimental work demonstrating that the correlated ground state of the pyrochlore material $Ce_2Sn_2O_7$ is a quantum liquid of magnetic octupoles, we establish a fundamentally new state of matter: a quantum ice of higher-rank multipoles. This discovery is ground-breaking for several reasons that are briefly summarized below [29,46].

States of matter that are primarily driven by multipolar interactions are rare, and the octupole ice is unique among those rarities in that it is highly frustrated and likely exhibits topological order. As stated at the outset, higher multipolar interactions are known or conjectured to be the key to understanding a wider range of 'hidden' orders in condensed-matter systems, from heavy-fermion materials to, potentially, high-transition-temperature superconductors. Bringing together this area of research with the intriguing world of 'ice physics', and quantum spin liquids more generally, holds the promise to open new perspectives.

The experimental results obtained here are consistent with the realization of a topological phase in three dimensions, known as the U(1) quantum spin liquid. In $Ce_2Sn_2O_7$ we have demonstrated that the relative contributions of dipole and octupole moments to the ground-state wavefunction depend on temperature, and also can be tuned by an applied magnetic field. These two control knobs on the 'dipole–octupole' quantum object encoded in the local ground-state wavefunction open up intriguing prospects with a view to explore the rich physics of these quantum many-body systems.

With the recent progress in the theoretical understanding of 'dipole-octupole' pyrochlores [47-50] and of quantum spin ice in general [37-39,51-52], the prospects for $Ce_2Sn_2O_7$ and related materials are very encouraging. It appears likely that these will conclude to a recognized identification of a three-dimensional quantum spin liquid, in line with the historical works having introduced the concept known as 'quantum spin ice' about 15 years ago [53], with the difference, however, that the ice manifold is constructed from the magnetic octupole components of a pseudo-spin [30-31]. Experimentally, there are great challenges ahead, such as resolving the octupole ice diffuse scattering in three-dimensional reciprocal space, as well as the spinon continuum using finer energy resolution, and also understanding the field-induced dipole orders.